\documentclass[journal]{IEEEtran}

\IEEEoverridecommandlockouts
\usepackage[linesnumbered, ruled,vlined]{algorithm2e}
\newtheorem{proposition}{Proposition}
\newtheorem{remark}{Remark}
\newtheorem{lemma}{Lemma}
\usepackage{cite}
\usepackage[utf8]{inputenc} 
\usepackage[T1]{fontenc}
\usepackage{amssymb,amsfonts}
\usepackage[cmex10]{amsmath}
\allowdisplaybreaks
\usepackage{graphicx}
\usepackage{xcolor}

\ifCLASSOPTIONcompsoc
    \usepackage[caption=false, font=normalsize, labelfont=sf, textfont=sf]{subfig}
\else
\usepackage[caption=false, font=footnotesize]{subfig}
\fi

\newtheorem{theorem}{Theorem}

\DeclareUnicodeCharacter{2212}{-}
\def\BibTeX{{\rm B\kern-.05em{\sc i\kern-.025em b}\kern-.08em
    T\kern-.1667em\lower.7ex\hbox{E}\kern-.125emX}}

\begin{document}
\title{Recursive Decoding of Reed-Muller Codes Starting With the Higher-Rate Constituent Code}

\author{Mikhail~Kamenev%
\thanks{This paper has been presented in part at the 2021 IEEE International Symposium on Information Theory \cite{SeqDec}.}%
\thanks{M. Kamenev was with the Moscow Research Center, Huawei Technologies Co., Ltd., Moscow, Russia. Email: mikhailkamenev92@gmail.com}}

\maketitle

\begin{abstract}
Recursive list decoding of Reed-Muller (RM) codes, with moderate list size, is known to approach maximum-likelihood (ML) performance of short length $(\leq 256)$ RM codes.
Recursive decoding employs the Plotkin construction to split the original code into two shorter RM codes with different rates. 
In contrast to the standard approach which decodes the lower-rate code first, the method in this paper decodes the higher-rate code first.
This modification enables an efficient permutation-based decoding technique, with permutations being selected on the fly from the automorphism group of the code using soft information from a channel.
Simulation results show that the error-rate performance of the proposed algorithms, enhanced by a permutation selection technique, is close to that of the automorphism-based recursive decoding algorithm with similar complexity for short RM codes, while our decoders perform better for longer RM codes.
In particular, it is demonstrated that the proposed algorithms achieve near-ML performance for short RM codes and for RM codes of length $2^m$ and order $m − 3$ with reasonable complexity.

\end{abstract}
\IEEEpeerreviewmaketitle

\section{Introduction}
\IEEEPARstart{B}{inary} Reed-Muller (RM) codes are a family of error-correcting codes of length $n = 2^m$ introduced by Muller \cite{Muller} in 1954.
Shortly after, Reed proposed a hard-input majority-logic decoder that achieves bounded distance decoding \cite{Reed}.
If a priori probabilities for the received bits are available, the performance of hard-input decoders can be improved using soft-input recursive decoders \cite{Litsyn, KabaRec, Bossert, recDec}.
A list version of recursive decoding with moderate list size approaches maximum-likelihood (ML) decoding performance for short length ($\leq 256$) RM codes\cite{Dumer}.
The permutation group of RM codes can be used to improve the performance of both recursive and recursive list decoding algorithms\cite{Dumer,RMPerm, PermGross, RMAutDec, PermGross2, FHTRMPerm}.
However, these decoders need a large list size or a large number of permutations to perform close to the ML decoder for codes of length larger than 256 \cite{Ivanov2}.

Several other soft-input decoding algorithms for RM codes have been proposed recently \cite{RMBP, MinYe, Pfister, Pfister2, SRPA}.
For instance, a recursive projection-aggregation algorithm demonstrates near-ML decoding performance for second-order RM codes \cite{MinYe, SRPA}, while a recursive puncturing–aggregation algorithm is suitable for high-rate RM codes decoding \cite{Pfister}.
Although these algorithms allow for parallel implementation, their computational complexity is high compared to the recursive list decoder.
Note that decoding algorithms aiming to improve the average-case running time have been considered in \cite{GSD, seq1, seq2}.

In \cite{PermGross2}, an algorithm that selects a good factor-graph permutation on the fly to enhance the performance of recursive list decoding has been proposed. 
As a result, the decoding algorithm proposed in \cite{PermGross2} requires a smaller list size compared to that of recursive list decoding with the same performance.
However, since the complexity of the factor-graph permutation selection scheme is high, the error-rate performance of the algorithm introduced in \cite{PermGross2} is close to that of recursive list decoding with a similar running time.

In this paper, we consider decoding algorithms that allow for a low-complexity permutation selection technique.
Similar to recursive decoding, our algorithms employ the Plotkin construction to get two shorter length RM codes, but decode a higher-rate constituent code first.
Although the error-rate performance of the proposed algorithms without permutations is limited, we demonstrate that a clever choice of permutations makes our algorithms competitive with the automorphism-based \cite{RMAutDec} recursive decoding \cite{recDec} that uses random permutations.
Namely, simulation results show that the proposed decoders outperform the automorphism-based recursive decoder with similar computational complexity for RM codes of length larger than 256, while the performance is nearly the same for shorter length codes.
Furthermore, we demonstrate that our algorithms achieve near-ML performance for codes of order $m - 3$ with reasonable complexity.

The rest of the paper is organized as follows.
Section II briefly introduces RM codes and the RM codes' automorphism group.
In Section III, we introduce a decoding algorithm for RM codes of order $m - 3$ and present a permutation selection technique.
This algorithm is generalized to decode arbitrary order RM codes in Section IV. 
Numerical results are presented in Section V.
We conclude the paper in Section VI.

\section{Preliminaries}
In the following, we use bold lower case letters to denote vectors and bold upper case letters to denote matrices.
We denote the $i$-th element of a vector $\mathbf{x}$ as $\mathbf{x}_i$ and we assume that indexing starts with zero.
We use $\oplus$ to denote a sum modulo 2.

Denote by $f\left(v \right) = f\left(v_0, \dotsc , v_{m-1}\right)$ a Boolean function of $m$ variables that is written in the algebraic normal form.
Let $\mathbf{f}$ be the vector of length $2^m$ containing values of $f$ at all of its $2^m$ arguments.
The binary RM code $\mathcal{R}\left(r, m\right)$ of order $r$ and length $n = 2^m$, $0\leq r \leq m$, is the set of all vectors $\mathbf{f}$, where $f\left(v \right)$ is a Boolean function of degree at most $r$.
The RM code $\mathcal{R}\left(r+1, m+1\right)$ can be represented in a recursive manner using $|\mathbf{u}|\mathbf{u} \oplus \mathbf{v}|$ construction, where $\mathbf{u} \in \mathcal{R}\left(r+1, m\right)$ and $\mathbf{v} \in \mathcal{R}\left(r, m\right)$  \cite[Sec.~13.3]{Sloan}.
Throughout the paper, we call $\mathcal{R}\left(r+1, m\right)$ the \textit{higher-rate constituent code}.

RM codes have the automorphism group, which is isomorphic to the general affine group $GA(m)$ \cite[Sec.~13.9]{Sloan}.
Recall that the automorphism group (or permutation group\footnote{
The automorphism group and the permutation automorphism group of the code are the same only for binary codes.
Since we consider only binary Reed-Muller codes, we use the terms "automorphism group" and "permutation group" interchangeably.
}) of a code contains permutations of the code positions that transform any codeword of the code to another or the same codeword.
For RM codes, these transformations can be expressed in terms of Boolean functions as follows: replace $f\left(v_0, \dotsc , v_{m - 1}\right)$ with 
\begin{equation*}
f\left(\bigoplus\limits_{j = 0}^{m-1} a_{0,j}v_j \oplus \mathbf{b}_0, \dotsc , \bigoplus\limits_{j = 0}^{m-1} a_{m - 1,j}v_j \oplus \mathbf{b}_{m-1}\right),
\end{equation*}
  where $\mathbf{A} = \left(a_{i,j}\right)$ is an invertible $m \times m$ binary matrix and $\mathbf{b}$ is a binary vector \cite[Sec.~13.9]{Sloan}.

Let $i \in \left\{0,1, \dots,2^m - 1\right\}$ be a code bit position and let $\mathbf{i} \in \left\{0,1\right\}^m$ be its binary representation.
Then, for a given invertible $m \times m$ binary matrix $\mathbf{A}$ and a vector $\mathbf{b}$, a permutation $\pi$ from the automorphism group of a code can be written as $\pi\left(i\right) = \sum_{k = 0}^{m - 1} 2^k \mathbf{j}_k$, where $\mathbf{j} = \mathbf{A}\mathbf{i} \oplus \mathbf{b}$ \cite{MinYeReview}.

\section{Decoding of High-Rate RM Codes}
In this section, we present a version of the algorithm for RM codes of length $n = 2^m$ and order $m-3$.
The key idea of the algorithm is similar to that of recursive decoding, i.e., to repeatedly split the code into two shorter codes until an RM code allowing for efficient ML decoding.
However, in contrast to recursive decoding, we propose to first decode a higher-rate constituent code.
Although the error-rate performance of this algorithm is much worse than that of recursive decoding, we demonstrate that this algorithm allows for a low-complexity permutation selection technique that significantly improves its performance.

\subsection{Description of the Algorithm}
Consider an RM code $\mathcal{R}\left(m-3, m\right)$ of length $n$.
Suppose that $\mathbf{c} \in \mathcal{R}\left(m-3, m\right)$ is transmitted over a binary-input additive white Gaussian noise (BI-AWGN) channel and let $\mathbf{x}$ be the received vector.
The log-likelihood ratio (LLR) vector $\mathbf{y}$ is then defined as
\begin{equation}
\mathbf{y}_i \triangleq \ln \frac{p\left(\mathbf{x}_i|\mathbf{c}_i = 0\right)}{p\left(\mathbf{x}_i|\mathbf{c}_i = 1\right)} = \frac{2\mathbf{x}_i}{\sigma^2},
\end{equation}
where $\sigma^2$ is the noise variance of the channel.

Let $\mathbf{y}^m = \mathbf{y}$.
Recall that a codeword $\mathbf{c} \in \mathcal{R}\left(m-3, m\right)$ can be written as $\mathbf{c} = |\mathbf{u}|\mathbf{u} \oplus \mathbf{v}|$, where $\mathbf{u} \in \mathcal{R}\left(m-3, m-1\right)$, $\mathbf{v} \in \mathcal{R}\left(m-4, m-1\right)$, and $|\mathbf{x}|\mathbf{x^\prime}|$ denotes the concatenation of vectors $\mathbf{x}$ and $\mathbf{x^\prime}$.
Note that $\mathbf{u}$ is a codeword of an extended Hamming code.
Thus, it is possible to use a low-complexity Chase II  decoding algorithm \cite{Chase} to decode the first half of the LLR vector $\mathbf{y}^m$.
Assume that the first half of the vector $\mathbf{y}^m$ is decoded correctly, i.e.,
the Chase decoder returns $\mathbf{u}$ as the hard output.
Then the LLR vector $\mathbf{y}^{m-1}$ corresponding to $\mathbf{v}$ can be easily obtained as $\mathbf{y}^{m-1}_i = \left(1-2\mathbf{u}_i\right)\mathbf{y}^m_{n/2 + i}$, $0 \leq i < n/2$.

\begin{figure}[t]
\centering
\includegraphics[scale=0.99]{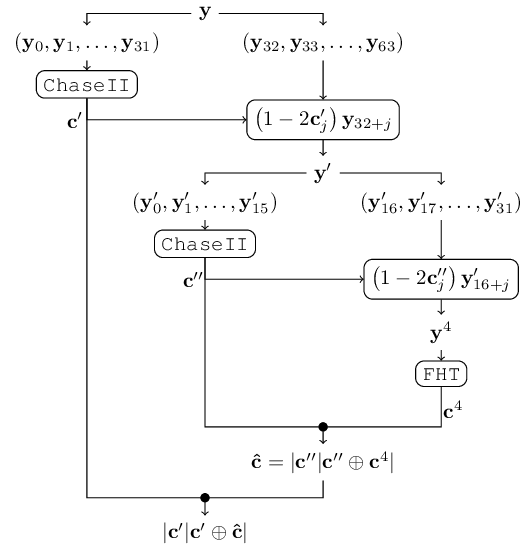}
\caption{Blockwise successive decoding of $\mathcal{R}\left(3, 6 \right)$. }
\label{BWS}
\end{figure}

Since $\mathbf{v} \in \mathcal{R}\left(m-4, m-1\right)$, it can be written as $\mathbf{v} = |\mathbf{u}^\prime|\mathbf{u}^\prime \oplus \mathbf{v}^\prime|$, where $\mathbf{u}^\prime \in \mathcal{R}\left(m-4, m-2\right)$ and $\mathbf{v}^\prime \in \mathcal{R}\left(m-5, m-2\right)$.
Thus, $\mathbf{y}^{m-1}$ can be processed in the same manner as $\mathbf{y}^m$.
This process continues until $\mathbf{y}^{4}$, i.e., a noisy codeword of $\mathcal{R}\left(1, 4\right)$, which is decoded using the fast Hadamard transform (FHT) \cite{Green}.
Throughout this paper, we refer to this decoding algorithm as a \textit{blockwise successive algorithm}\footnote{In \cite{SeqDec}, this algorithm is called sequential decoding. However, in the literature, the term sequential decoding is usually associated with another decoding technique (for instance, see \cite{stolte2000sequential}). Therefore, we use a different name for the proposed algorithm in this paper.}. 
Blockwise successive decoding of $\mathcal{R}\left(3, 6 \right)$ is illustrated in Fig. \ref{BWS}.

\begin{remark}
There is an efficient bit-wise maximum a posteriori decoder of the extended Hamming codes  that has computational complexity $\mathcal{O}\left(n\log n\right)$ \cite{LitsynMAP}.
Unfortunately, the output of this decoder is not necessarily a codeword of the extended Hamming code.
It is the main reason why blockwise successive decoding uses the Chase algorithm to decode an extended Hamming code.
\end{remark}

The formal description of the blockwise successive algorithm is given in Algorithm \ref{SDAlg}.
Note that this algorithm uses the Chase II algorithm with $m$ unreliable bits for decoding of length $2^m$ extended Hamming code.
We consider the following implementation of the Chase II algorithm in Algorithm \ref{SDAlg}.
Suppose that $\mathbf{y}_{i_0}, \mathbf{y}_{i_1}, \dots \mathbf{y}_{i_{m-1}}$ are $m$ LLRs with the smallest absolute values.
The Chase II algorithm enumerates all $2^m$ possible hard output for codeword bits corresponding to these unreliable LLRs.
The rest of hard values are assigned based on the sign of the LLR values.
Then, $2^m$ hard vectors are decoded using syndrome decoding of extended Hamming code.
Namely, if the syndrome equals a column in the parity-check matrix of the code, then a bit corresponding to this column is flipped.
Otherwise, the failure is declared for the given hard pattern of length $m$.
Consequently, this procedure can generate at most $2^m$ codewords of the extended Hamming code.
Finally, the algorithm computes a correlation discrepancy \cite[Sec.~10.1]{shulin} for each of these codewords and returns the codeword with the smallest metric value.

\begin{algorithm}[t]
\DontPrintSemicolon
\SetAlgoLined
\KwIn{An LLR vector $\mathbf{y}$ of length $n = 2^m$}
\KwOut{A decoded codeword $\mathbf{c}$}
Let $\mathbf{c}$ be zero vector of length $n$ \;
\For{$l = m-1, m-2, \dots, 4$}{
	$\mathbf{\hat{y}} \gets \left(\mathbf{y}_{2^m - 2^{l + 1}}, \mathbf{y}_{2^m - 2^{l + 1} + 1}, \dots , \mathbf{y}_{2^m - 2^l - 1}\right)$\;
	$\mathbf{\hat{c}} \gets \texttt{ChaseII}\left(\mathbf{\hat{y}}\right)$ \tcp*{\texttt{ChaseII} function decodes an input LLR vector of length $2^l$ using the Chase II algorithm with $l$ unreliable bits \cite{Chase}} 
	\For {$i = 0, 1, \dots 2^l - 1$} {
		$\mathbf{y}_{2^m - 2^l + i} \gets \left(1 - 2\mathbf{\hat{c}}_i\right)\mathbf{y}_{2^m - 2^l + i}$\;
		$\mathbf{c}_{2^m - 2^{l+1} + i} \gets \mathbf{c}_{2^m - 2^{l+1} + i} \oplus \mathbf{\hat{c}}_i$\;
	   $\mathbf{c}_{2^m - 2^l + i} \gets \mathbf{c}_{2^m - 2^l + i} \oplus \mathbf{\hat{c}}_i$\;
	}
}
$\mathbf{y}^4 \gets \left(\mathbf{y}_{2^m - 16}, \mathbf{y}_{2^m - 15}, \dots , \mathbf{y}_{2^m - 1}\right)$\;
$\mathbf{c}^4 \gets \texttt{FHTDec}\left(\mathbf{y}^4\right)$  \tcp*{\texttt{FHTDec} function performs ML decoding of an input LLR vector using the FHT-based algorithm \cite{Green}}
\For {$i = 0, 1, \dots 15$} {
$\mathbf{c}_{2^m - 16 + i} \gets \mathbf{c}_{2^m - 16 + i} \oplus \mathbf{c}^4_i$ \;
}
 \Return{$\mathbf{c}$}
 \caption{The \texttt{BWSDec} decoding function}
 \label{SDAlg}
\end{algorithm}

\begin{lemma}
Algorithm \ref{SDAlg} takes $\mathcal{O}\left(n\log n\right)$ time.
\end{lemma}
\begin{IEEEproof}
First, we show that the running time of all \texttt{ChaseII} function calls used in Algorithm \ref{SDAlg} is $\mathcal{O}\left(n\log n\right)$.
Consider the \texttt{ChaseII} function and assume that its input is a vector $\mathbf{y}^l$ of length $n^\prime = 2^l$.
In the beginning, this function uses sorting, which has running time $\mathcal{O}\left(n^\prime\log n^\prime\right)$, to find unreliable LLRs.
Then, the algorithm runs syndrome decoding $n^\prime$ times.
The computational complexity of this procedure is optimized as follows.
Suppose that the hard decision is made based on the LLR vector and the syndrome $\mathbf{s}$ is calculated. 
We assume that, in the case of an extended Hamming code, the syndrome computation can be done using a summation of integer numbers modulo 2.
Therefore, the running time of the syndrome's calculation is $\mathcal{O}\left(n^\prime\right)$.
If $\mathbf{s}$ is known, then the calculation of syndrome for each hard pattern of length $l$ requires at most $l$ summations modulo 2.
Consequently, the computational complexity of the syndromes' calculation is $\mathcal{O}\left(n^\prime\log n^\prime\right)$.
Since an integer representation $s$ of the syndrome $\mathbf{s}$ is less than $2n^\prime$, one can use an array of length $2n^\prime$ to find a position of error bit with the running time $\mathcal{O}\left(1\right)$.
As a result, the running time of this part of the function \texttt{ChaseII} is $\mathcal{O}\left(n^\prime\log n^\prime\right)$.

The correlation discrepancy of a codeword $\mathbf{c}$ is calculated as $\sum_{i \in \mathcal{I}}\left|\mathbf{y}^l_i\right|$, where  $\mathcal{I} = \left\lbrace i: \mathrm{sign}\left(\mathbf{y}^l_i \right) \neq \left(1 - 2\mathbf{c}_i \right) \right\rbrace$ \cite[Sec.~10.1]{shulin}.
Observe that the algorithm calculates the correlation discrepancy for codewords such that $\left|\mathcal{I} \right| \leq l + 1$.
Consequently, all metric values are calculated in $\mathcal{O}\left(n^\prime\log n^\prime\right)$.
Therefore, the running time of \texttt{ChaseII}$\left(\mathbf{y}^l \right)$ is $\mathcal{O}\left(n^\prime\log n^\prime\right)$.
Note that the function \texttt{ChaseII} is applied to vectors of lengths $2^5, 2^6, \dots, 2^{m-1}$.
Hence, the total running time of this function in Algorithm \ref{SDAlg} is  $\mathcal{O}\left(n\log n\right)$. 

Since the operations in lines 6 -- 8 of Algorithm \ref{SDAlg} have linear computational complexity, the running time of the for loop in lines 2 -- 10 of Algorithm \ref{SDAlg} is $\mathcal{O}\left(n\log n\right)$.
Note that lines 11 -- 15 of Algorithm \ref{SDAlg} deal with a vector of length 16.
Therefore, this part of the algorithm takes constant running time. 
It follows that the running time of Algorithm \ref{SDAlg} is indeed $\mathcal{O}\left(n\log n\right)$.
\end{IEEEproof}

\begin{lemma}
The space complexity of a sequential implementation of Algorithm \ref{SDAlg} is $\mathcal{O}\left(n\right)$.
\end{lemma}
\begin{IEEEproof}
Consider the function \texttt{ChaseII} and assume that its input is a vector $\mathbf{y}^l$ of length $n^\prime = 2^l$. 
Sorting used in this function can be implemented in place \cite[Part~II]{Cormen}. 
Thus, the total memory required for sorting is $\mathcal{O}\left(n^\prime\right)$.

The syndrome decoding used in the function \texttt{ChaseII} needs to store at most $l+1$ positions of the received LLRs with the incorrect sign, the value of the correlation discrepancy, and an array of size $2n^\prime$ that maps the syndrome value to the position of incorrect bit.
Also, the algorithm needs to store the error positions corresponding to the codeword with the smallest correlation discrepancy and the metric of this codeword.
Since we consider a sequential implementation of this algorithm, the space complexity of syndrome decoding is $\mathcal{O}\left(n^\prime\right)$.
Consequently, the function \texttt{ChaseII} has linear space complexity.

Since the function \texttt{FHTDec} and the for loop in lines 13--15 deal with a vector of constant length, it follows that lines  11--15 use a constant amount of memory. 
Thus, the space complexity of Algorithm \ref{SDAlg} is $\mathcal{O}\left(n\right)$.
\end{IEEEproof}

\begin{remark}
Recursive decoding of an arbitrary order RM code takes $\mathcal{O}\left(n\log n\right)$ time.
However, since the complexity of recursive decoding is upper bounded by $3n \cdot \min\left\{r, m - r \right\} + n\left(m - r \right) + n$ \cite{recDec}, the running time of recursive decoding for RM codes of order $m - 3$ grows linearly with the code length and, as a consequence, grows slower compared to blockwise successive decoding.
The space requirements of these algorithms are similar.
\end{remark}

Note that the output of Algorithm \ref{SDAlg} is a codeword of $\mathcal{R}\left(m - 3, m\right)$.
We prove it by the induction on $m$.
Consider the base case of $m = 4$.
In this case, the output of the algorithm equals the output of the FHT-based decoder of the first-order RM codes, and the claim holds.
Assume that the claim holds for all $\mathcal{R}\left(m - 3, m\right)$ and we prove it for $\mathcal{R}\left(m - 2, m + 1\right)$.
Consider the for-loop in lines 2 -- 10 of Algorithm \ref{SDAlg}.
In the first iteration, the Chase decoder is used to get a codeword of $\mathcal{R}\left(m - 2, m\right)$ and it is assigned to the first and the second halves of the vector $\mathbf{c}$.
Then, the algorithm processes the second half of the LLR vector $\mathbf{y}$ and, by the induction hypothesis, returns a codeword of $\mathcal{R}\left(m - 3, m\right)$.
This codeword is added to the second half of the vector $\mathbf{c}$ (see lines 7--8 and line 14).
Therefore, the output of Algorithm \ref{SDAlg} can be written as $\mathbf{c} = |\mathbf{u}|\mathbf{u} \oplus \mathbf{v}|$, where $\mathbf{u} \in \mathcal{R}\left(m-2, m\right)$, $\mathbf{v} \in \mathcal{R}\left(m-3, m\right)$.
Thus, $\mathbf{c} \in \mathcal{R}\left(m - 2, m+1\right)$.

\subsection{Permutation-Based Blockwise Successive Decoding}
Consider blockwise successive decoding of $\mathcal{R}\left(m-3, m\right)$.
Observe that $\mathcal{R}\left(m-3, m-1\right)$ has a higher rate than $\mathcal{R}\left(m-3, m\right)$.
As a consequence, $\mathcal{R}\left(m-3, m-1\right)$ has a higher block error probability under ML decoding than $\mathcal{R}\left(m-3, m\right)$.
Since the blockwise successive algorithm uses the original LLR vector for decoding of  $\mathcal{R}\left(m-3, m-1\right)$, the block error probability of $\mathcal{R}\left(m-3, m\right)$ under blockwise successive decoding is lower bounded by the block error probability of the extended Hamming code under ML decoding.
Therefore, the performance of the blockwise successive decoder is very poor.

The performance of blockwise successive decoding can be enhanced by permutations from the automorphism group of the code.
For instance, different permuted LLR vectors are decoded using the blockwise successive algorithm and then the output codewords are de-interleaved.
The output of this algorithm is a codeword with the best metric.
A similar approach has been used for recursive decoding \cite{RMPerm, RMAutDec, FHTRMPerm} and it allows to improve the performance of the recursive algorithm significantly.
However, we found that permutations for blockwise successive decoding can be selected based on soft information from a channel.
It results in a better performance in comparison with the case of random permutations.

In this subsection, we present an efficient algorithm for the selection of permutations from the automorphism group of the code for blockwise successive decoding.
This algorithm aims to find a permutation that moves reliable LLRs, i.e., LLRs with large absolute values, to the first half of the vector, while unreliable LLRs, i.e., LLRs with small absolute values, are moved as close to the end of the vector as possible.
The intuition behind why such permutations improve the error-rate performance of blockwise successive decoding is that the resulting permuted vector contains a small fraction of unreliable LLRs in the first half of the vector and, as a consequence, it increases the probability of successful decoding of the $\mathcal{R}\left(m-3, m-1\right)$ constituent code.

Consider a \texttt{PermTransform} function presented in Algorithm \ref{PermTransform}.
This function takes as input a random permutation $\pi$ of length $n = 2^m$ and transforms it into a permutation $\bar{\pi}$ from the automorphism group of the code.
In the next proposition, we prove the correctness of this function.

\begin{algorithm}
\DontPrintSemicolon
\SetAlgoLined
\KwIn{A random permutation $\pi$ of length $n$}
\KwOut{A permutation $\bar{\pi}$ from the automorphism group of length $n$ RM codes}
 Let $\hat{\pi}$ be a permutation of length $n$\;
 $\hat{\pi}\left(0\right) \gets \pi\left(0\right)$ \;
 Let $\mathbf{x}$ be zero vector of length $n$ \;
 $\mathbf{x}_{\hat{\pi}\left(0\right)} \gets 1$\;
 $i \gets 1$\;
 \For{$l = 0, 1, \dots, \log_2{\left(n\right)} - 1$}{
 	
 	\While{$\mathbf{x}_{\pi\left(i\right)} = 1$} {
 		$i \gets i + 1$\;
 	}
 	$\hat{\pi}\left(2^l\right) \gets \pi\left(i \right)$\;
 	$\mathbf{x}_{\hat{\pi}\left(2^l\right)} \gets 1$\;
 	$i \gets i + 1$\;
 	\For{$t = 2^l + 1, 2^l + 2, \dots, 2^{l+1} - 1$}{
 		$\hat{\pi}\left(t\right) \gets \hat{\pi}\left(t - 2^l \right) \oplus \hat{\pi}\left(2^l\right) \oplus \hat{\pi}\left(0\right)$ \;
 		$\mathbf{x}_{\hat{\pi}\left(t\right)} \gets 1$\;
 	}
 }
 Let $\bar{\pi}$ be a permutation of length $n$ such that $\bar{\pi}\left(i\right) = \hat{\pi}\left(n - i - 1\right)$, $0 \leq i \leq n - 1$\;
 \Return{$\bar{\pi}$}
 \caption{The \texttt{PermTransform} function for the permutation-based blockwise successive decoding algorithm}
 \label{PermTransform}
\end{algorithm}

\begin{proposition}
The output of the \texttt{PermTransform} function is a permutation from the automorphism group of RM codes.
\end{proposition}
\begin{IEEEproof}
We first show that the auxiliary permutation $\hat{\pi}$ used in Algorithm \ref{PermTransform} is from the automorphism group of the code.
Recall that a permutation from the automorphism group of the code can be defined using matrix multiplication as 
$\hat{\pi}\left(i\right) = \sum_{k = 0}^{m - 1} 2^k \mathbf{j}_k$, where $\mathbf{j} = \mathbf{A}\mathbf{i} \oplus \mathbf{b}$, $\mathbf{i}$ is a binary representation of index $i$, $\mathbf{A} = \left(a_{r,c}\right)$ is an invertible $m \times m$ binary matrix, and $\mathbf{b}$ is a binary vector \cite{MinYeReview}.
Observe that $\hat{\pi}\left(0\right) = \sum_{k = 0}^{m - 1} 2^k \mathbf{b}_k$.
Therefore, the \texttt{PermTransform} function chooses as $\mathbf{b}$ a binary representation of $\pi\left(0\right)$ (line 2).
If a binary representation of an index $i$ contains only one non-zero entry, i.e.,
indices $i = 2^l$, $l \in \left\lbrace 0, \dots, m - 1\right\rbrace$, then $\hat{\pi}\left(i\right) = \sum_{k = 0}^{m - 1} 2^k\left(a_{k, l} \oplus  \mathbf{b}_k\right)$.
Consequently, $\hat{\pi}\left(2^0\right), \hat{\pi}\left(2^1\right), \dots , \hat{\pi}\left(2^{m-1}\right)$ define the columns of the matrix $\mathbf{A}$ plus the vector $\mathbf{b}$ (line 10).
The while loop in lines 7 -- 9 is used to guarantee that the matrix $\mathbf{A}$ corresponding to the permutation $\hat{\pi}$ is invertible.

It remains to show that $\hat{\pi}\left(i\right)$ calculated in lines 13--16 equals $\sum_{k = 0}^{m - 1} 2^k \mathbf{j}_k$, where $\mathbf{j} = \mathbf{A}\mathbf{i} \oplus \mathbf{b}$.
We prove it by induction on $l$ used in for loop in lines 6 -- 17.
Observe that the claim holds for the base case of $l = 1$ (it is the smallest $l$, for which the algorithm enters the loop in lines 13--16).
Indeed,  
\begin{equation}
\begin{aligned}
\hat{\pi}\left(3\right) &= \sum\limits_{k = 0}^{m - 1}2^k\left(a_{k, 0} \oplus a_{k,1} \oplus \mathbf{b}_k \right)  \\
& = \left(\sum\limits_{k = 0}^{m - 1}2^k\left(a_{k, 0} \oplus \mathbf{b}_k \right) \right) \\
& \oplus \left(\sum\limits_{k = 0}^{m - 1}2^k\left(a_{k,1} \oplus \mathbf{b}_k \right) \right) \oplus \sum\limits_{k = 0}^{m - 1}2^k \mathbf{b}_k \\
& = \hat{\pi}\left(1\right) \oplus \hat{\pi}\left(2\right)  \oplus \hat{\pi}\left(0\right).
\end{aligned}
\end{equation}
Let us assume that the claim holds for $l$ and we prove it for $l + 1$.
Observe that $\hat{\pi}\left(i\right) = \sum_{k = 0}^{m - 1} 2^k \mathbf{j}_k, i \leq 2^l$, where $\mathbf{j} = \mathbf{A}\mathbf{i} \oplus \mathbf{b}$.
Consider an index $t$, $2^l < t < 2^{l+1}$.
Denote by $\mathbf{\hat{l}}$ a binary representation of $2^l$ and denote by $\mathbf{\hat{l}^t}$ a binary representation of $t - 2^l$.
Let $\mathbf{\hat{z}} = \mathbf{A}\mathbf{\hat{l}}$, let $\mathbf{\hat{z}^t} = \mathbf{A}\mathbf{\hat{l}^t}$, and let $\mathbf{z} = \mathbf{A}\mathbf{t}$.
Since $t \in \left\lbrace 2^l + 1, \dots,  2^{l+1} - 1 \right\rbrace$, it follows that $t = \left(t - 2^l\right) \oplus 2^l$.
Consequently,
\begin{equation}
\begin{aligned}
\hat{\pi}\left(t\right) &= \sum\limits_{k = 0}^{m - 1}2^k\left(\mathbf{z}_k \oplus \mathbf{b}_k \right) = \sum\limits_{k = 0}^{m - 1}2^k\left(\mathbf{\hat{z}}_k \oplus \mathbf{\hat{z}^t}_k\oplus \mathbf{b}_k \right) \\
&= \left(\sum\limits_{k = 0}^{m - 1}2^k\left(\mathbf{\hat{z}}_k \oplus \mathbf{b}_k \right) \right) \oplus \left(\sum\limits_{k = 0}^{m - 1}2^k\left(\mathbf{\hat{z}^t}_k  \oplus \mathbf{b}_k \right) \right) \\ 
& \oplus  \sum\limits_{k = 0}^{m - 1}2^k\mathbf{b}_k = \hat{\pi}\left(2^l \right) \oplus  \hat{\pi}\left(t - 2^l \right) \oplus \hat{\pi}\left(0\right).
\end{aligned}
\end{equation}
This establishes the inductive step and completes the proof that $\hat{\pi}$ is a permutation from the automorphism group of RM codes.

Note that the permutation $\pi^\prime$ of length $n$, $\pi^\prime\left(i\right) = n - 1 - i$, is in the automorphism group of the code ($\pi^\prime\left(i\right) = \sum_{k = 0}^{m - 1} 2^k \left(\mathbf{i}_k \oplus 1\right)$, where $\mathbf{i}$ is a binary representation of $i$).
Consequently, $\hat{\pi} \circ \pi^\prime$ is in the automorphism group of the code.
Since the permutation $\bar{\pi}$ returned by the function \texttt{PermTransform} can be written as $\hat{\pi} \circ \pi^\prime$, the function \texttt{PermTransform} indeed returns a permutation from the automorphism group of the code.
\end{IEEEproof}

Consider an LLR vector $\mathbf{y} = \left(y_0, y_1, \dots, y_{n-1}\right)$, $n = 2^m$.
Let $\pi$ be a permutation such that the vector of LLR's absolute values  $\left(|y_{\pi\left(0\right)}|, |y_{\pi\left(1\right)}|, \dots, |y_{\pi\left(n-1\right)}| \right)$ is ordered in the ascending order and denote by $\bar{\pi}$ the result of $\texttt{PermTransform}\left(\pi\right)$.
Observe that for any $j \in \left\lbrace 0, \dots, 4 \right\rbrace$, there exists $i \in \left\lbrace n-16,n-15 \dots, n-1 \right\rbrace$ such that $\bar{\pi}\left(i\right) = \pi\left( j\right)$.
Therefore, there are at least 5 unreliable LLRs among the last 16 elements of the vector $\left(y_{\bar{\pi}\left(0\right)}, y_{\bar{\pi}\left(1\right)}, \dots, y_{\bar{\pi}\left(n-1\right)} \right)$.
Thus, if the blockwise successive algorithm decodes this vector, then at least 5 unreliable LLRs will be processed by FHT in the last step of the decoding algorithm.

\begin{remark}
The idea of dividing the bit positions into two disjoint sets based on their reliabilities has been used in \cite{RMBP} to generate the rows of an overcomplete parity-check matrix tailored to belief propagation decoding of the received sequence.
Specifically, the algorithm proposed in \cite{RMBP} generates an $\left(r+1\right) \times m$ matrix and then matrix multiplication is used to find $2^{r+1}$ non-zero positions of a minimum-weight parity check.
The matrix is selected in such a way that the resulting parity check contains at least one unreliable position and $r+1$ reliable ones.
In \cite{permSearch}, a permutation selection approach similar to Algorithm \ref{PermTransform} has been proposed.
Namely, the binary expansion of indices corresponding to the least reliable LLRs is used in \cite{permSearch} to generate an invertible matrix.
Then matrix multiplication is used to find the permutation  associated with the matrix.
In contrast to the method proposed in \cite{permSearch}, Algorithm \ref{PermTransform} does not generate an invertible matrix and directly returns a permutation from the automorphism group of the code.
\end{remark}

\begin{lemma}
The running time and the space complexity of Algorithm \ref{PermTransform} are $\mathcal{O}\left(n\right)$.
\end{lemma}
\begin{IEEEproof}
Algorithm \ref{PermTransform} uses two summations modulo 2 to calculate the majority of $\hat{\pi}\left(t\right)$ and simple assigning for the others.
Consequently, the complexity of the $\hat{\pi}$ calculation is  $\mathcal{O}\left(n \right)$.
Note that this function also uses a simple check in lines 7--9.
The complexity of this check in the worst case is also $\mathcal{O}\left(n \right)$.
Since the permutation $\bar{\pi}$ is derived from $\hat{\pi}$ by reversing the order, the running time of Algorithm \ref{PermTransform} is $\mathcal{O}\left(n \right)$.

Algorithm \ref{PermTransform} allocates memory to store the output permutation $\bar{\pi}$, the auxiliary permutation $\hat{\pi}$, and the temporary array $\mathbf{x}$ of size $n$.
Thus, the space complexity of Algorithm \ref{PermTransform} is $\mathcal{O}\left(n \right)$.
\end{IEEEproof} 

\begin{figure}[!t]
\centering
\includegraphics[scale=0.99]{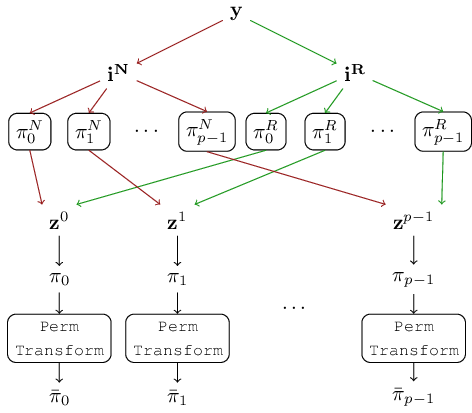}
\caption{The permutation selection for the permutation-based blockwise successive decoding}
\label{PBSD1}
\end{figure}

We now introduce the permutation-based blockwise successive decoding algorithm. 
The formal description of this algorithm is presented in Algorithm \ref{PBSDAlg}.
First, we describe an algorithm for permutation selection, which is illustrated in Fig. \ref{PBSD1}.
In the beginning, sorting is used to find indices of $l$ LLRs with the smallest absolute values $\mathbf{i^N} = \left(i^N_0, i^N_1, \dots, i^N_{l-1} \right)$.
Denote by $\mathbf{i^R} = \left(i^R_0, i^R_1, \dots, i^R_{n-l-1} \right)$ a vector with the remaining $n - l$ indices.
Then, $\mathbf{i^N}$ and $\mathbf{i^R}$ are permuted using $p$ pairs of random  permutations $\pi^N_j$ and $\pi^R_j$, $0 \leq j < p$.
Let $\mathbf{z}^j = \left(i^N_{\pi^N_j\left(0\right)}, i^N_{\pi^N_j\left(1\right)}, \dots, i^N_{\pi^N_j\left(l-1\right)}, i^R_{\pi^R_j\left(0\right)}, i^R_{\pi^R_j\left(1\right)}, \dots, i^R_{\pi^R_j\left(n - l-1\right)}\right)$.
We use $\mathbf{z}^j$ to define a permutation $\pi_j$ as $\pi_j\left(k \right) = \mathbf{z}^j_k$, $0 \leq k \leq n - 1$.
These permutations are transformed into permutations from the automorphism group of the code using the \texttt{PermTransform} function.
Denote the transformed permutations as $\bar{\pi}_j$, $0 \leq j < p$.

\begin{figure}[t]
\centering
\includegraphics[scale=0.99]{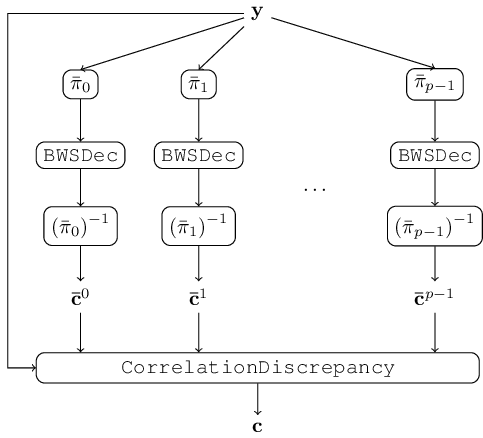}
\caption{The permutation-based blockwise successive decoding algorithm with permutations $\bar{\pi}_0, \bar{\pi}_1, \dots \bar{\pi}_{p-1}$. }
\label{PBSD2}
\end{figure}

The second part of the permutation-based algorithm is illustrated in Fig. \ref{PBSD2}.
At this stage, the permutations $\bar{\pi}_j$ are used in a manner similar to the permutation decoder proposed in \cite{RMPerm}.
Namely, the LLR vector is permuted using the $\bar{\pi}_j$ permutations and each permuted version of the LLR vector is decoded using the blockwise successive algorithm.
Finally, the output of each blockwise successive decoder is de-interleaved and the algorithm returns a codeword $\mathbf{c}$ with the smallest correlation discrepancy.

Now we prove that the proposed permutation-based algorithm has computational complexity $\mathcal{O}\left(pn \log n \right)$ for both sequential and parallel implementations.
By the parallel implementation of this algorithm, we mean an implementation that parallelizes the loop in lines 5 -- 18 of Algorithm \ref{PBSDAlg}.

\begin{algorithm}[!t]
\DontPrintSemicolon
\SetAlgoLined
\KwIn{An LLR vector $\mathbf{y}$ of length $n = 2^m$, a number of unreliable LLRs $l$, a number of permutations $p$}
\KwOut{A decoded codeword $\mathbf{c}$}
Let $\mathbf{i^N}$ be an array of length $l$ that contains indices of $l$ least reliable LLRs in  $\mathbf{y}$ \;
Let $\mathbf{i^R}$ be an array of length $n-l$ that contains indices of $n-l$ most reliable LLRs in  $\mathbf{y}$ \;
Let $\mathbf{c}$ be zero vector of length $n$ \;
$M \gets \infty$ \;
\For{$k = 0, 1, \dots, p - 1$} {
Let $\pi\left(\mathbf{i^N}\right)$ be a randomly permuted copy of the vector $\mathbf{i^N}$\;
Let $\pi\left(\mathbf{i^R}\right)$ be a randomly permuted copy of the vector $\mathbf{i^R}$\;
Let $\mathbf{z}$ be a concatenation of $\pi\left(\mathbf{i^N}\right)$ and $\pi\left(\mathbf{i^R}\right)$, $\mathbf{z} = \left|\pi\left(\mathbf{i^N}\right)  | \pi\left(\mathbf{i^R}\right) \right|$\;
Let $\pi$ be a permutation of length $n$ such that $\pi\left(i \right) = \mathbf{z}_i$, $0 \leq i \leq n - 1$ \;
$\bar{\pi} \gets \texttt{PermTransform}\left(\pi\right)$ \;
$\mathbf{y}^{\prime} \gets \left(y_{\bar{\pi}\left(0\right)}, y_{\bar{\pi}\left(1\right)}, \dots, y_{\bar{\pi}\left(n-1\right)} \right)$ \;
$\mathbf{\bar{c}}^\prime \gets \texttt{BWSDec}\left(\mathbf{y}^{\prime}\right)$\;
Let $\mathbf{\bar{c}}$ be a vector of size $n$, $\mathbf{\bar{c}}_{\bar{\pi}\left(i\right)} = \mathbf{\bar{c}}^\prime_i$, $0 \leq i \leq n - 1$\;
Let $\bar{M}$ be the correlation discrepancy of the codeword $\mathbf{\bar{c}}$\;
\If {$\bar{M} < M$} {
$\mathbf{c} \gets \mathbf{\bar{c}}$, $M \gets \bar{M}$ \;
}
}
 \Return{$\mathbf{c}$}
 \caption{The \texttt{PermBWSDec} decoding function}
 \label{PBSDAlg}
\end{algorithm}

\begin{theorem}
The running time of the permutation-based blockwise successive decoding algorithm is $\mathcal{O}\left(pn \log n \right)$ for both sequential and parallel implementations.
\end{theorem}
\begin{IEEEproof}
Consider the running time of each part of Algorithm \ref{PBSDAlg}.
In lines 1 -- 2,
Algorithm \ref{PBSDAlg} creates an array of length $l$ that stores indices of unreliable LLRs and an array of length $n - l$ that stores indices of reliable LLRs.
One can use sorting to create these arrays.
Consequently, the complexity of this operation is $\mathcal{O} \left( n \log n\right)$.
Next, the algorithm permutes these arrays using $p$ pairs of random permutations.
The complexity of this operation is $\mathcal{O} \left( pn\right)$.

Then, the algorithm uses the \texttt{PermTransform} function to create $p$ permutations from the automorphism group of the code.
Since the running time of this function is $\mathcal{O}\left(n\right)$, the total computational complexity of permutation selection is $\mathcal{O}\left(p n\right)$.

Next, the algorithm permutes the LLR vector, decodes the permuted vector using Algorithm \ref{SDAlg}, and de-interleaves the blockwise successive decoding output.
The total computational complexity of these operations is $\mathcal{O}\left(pn \log n \right)$.
Finally, the correlation discrepancy is computed for each output codeword and a codeword with the smallest metric is returned.
The correlation discrepancy for all codeword is calculated in $\mathcal{O}\left(pn\right)$.

We can see that the computational complexity of Algorithm \ref{PBSDAlg} is dominated by the complexity of the blockwise successive algorithm.
Therefore, the running time of the permutation-based blockwise successive decoding algorithm is $\mathcal{O}\left(pn \log n \right)$.
Observe that the parallel implementation of this algorithm does not require any additional calculations.
This completes the proof of the theorem.
\end{IEEEproof}

We now consider the space complexity of permutation-based blockwise successive decoding.
Sorting that is used for the calculation of arrays $\mathbf{i^N}$ and $\mathbf{i^R}$ can be implemented in place \cite[Part~II]{Cormen}.
Consequently, the calculation of these arrays requires $\mathcal{O}\left(n \right)$ space.
Observe that, in each iteration of the for loop in lines 5 -- 18, the algorithm allocates several arrays of length at most $n$ and uses the functions \texttt{PermTransform} and \texttt{BWSDec} that have a linear space complexity.
Therefore, the space complexity of the sequential implementation of Algorithm \ref{PBSDAlg} is $\mathcal{O}\left(n \right)$, while the parallel implementation takes $\mathcal{O}\left(pn \right)$ space.

\section{Decoding of Arbitrary Order RM Codes}
In this section, we propose a generalized blockwise successive decoding algorithm that can be applied to arbitrary order RM codes.
We observed that, even though the higher-rate constituent code is decoded by a sub-optimal algorithm, e.g., automorphism-based recursive decoding, it is possible to find a permutation that leads to correct decoding of this constituent code.
Thus, we propose to use two sub-optimal algorithms to decode shorter length constituent codes, with a higher-rate one being decoded first.

Another feature of the generalized algorithm is that it decodes different higher-rate constituent codes only once.
Observe that different blockwise successive decoders in Algorithm \ref{PBSDAlg} may process the same constituent codes in the first iteration of for-loop in lines 2--10 of Algorithm \ref{SDAlg}.
Consequently, it is possible to decrease computational complexity by processing these codes only once.
A naive implementation of this approach checks whether there are permutations $\bar{\pi}_i, 0 \leq i < p$, that result in the same constituent code in the first iteration of blockwise successive decoding and processes them only once.
However, this approach does not guarantee that such permutations exist.
Thus, it only allows decreasing the average computational complexity of the algorithm.
To address this issue, we propose a method that selects $p$ LLR vectors corresponding to different higher-rate constituent codes using puncturing.

Consider a codeword of $\mathcal{R}\left(r, m\right)$.
Observe that there are $2n - 2$, $n = 2^m$, ways to puncture bits in this codeword to get a vector from $\mathcal{R}\left(r, m-1\right)$.
Indeed, any codeword of an RM code $\mathcal{R}\left(r, m\right)$ comes from a polynomial 
$f\left(v_0, \dotsc , v_{m-1}\right) = g\left(v_0, \dotsc , v_{m - 2}\right) \oplus v_{m-1}h\left(v_0, \dotsc , v_{m - 2}\right)$, where $\deg\left(g\right) \leq r$ and $\deg\left(h\right) \leq r - 1$ \cite[Sec.~13.3]{Sloan}.
Note that the vector corresponding to a polynomial $v_{m - 1}$ has Hamming weight of $n/2$.
Therefore, puncturing of $n/2$ bits can be done in the following way: remove the codeword bits corresponding to non-zero values of $v_{m - 1}$.
Consequently, a codeword of the punctured code comes from the polynomial $g$.
Since $\deg\left(g\right) \leq r$ and the length of the punctured code equals $2^{m-1}$, the punctured code is $\mathcal{R}\left(r, m-1\right)$.
Moreover, it is possible to use permutations from the code's automorphism group to change the puncturing pattern.
Namely, one can replace $f\left(v_0, \dotsc , v_{m-1}\right)$ with $f\left(\bigoplus a_{0,j}v_j \oplus \mathbf{b}_0, \dotsc , \bigoplus a_{m - 1,j}v_j \oplus \mathbf{b}_{m-1}\right)$, where $\mathbf{A} = \left(a_{i,j}\right)$ is an invertible $m \times m$ binary matrix and $\mathbf{b}$ is a binary vector.
As a result, the puncturing patterns are defined by the polynomials $\bigoplus a_{m - 1,j}v_j \oplus \mathbf{b}_{m-1}$.
It is easy to verify that there are $2n - 2$ such polynomials.
Note that all codewords of the first-order RM code $\mathcal{R}\left(1, m\right)$ with Hamming weight $n/2$ come from these polynomials.

\begin{figure}[t]
\centering
\includegraphics[scale=0.99]{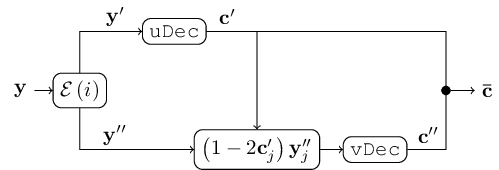}
\caption{Decoding of an LLR vector $\mathbf{y}$ by the generalized blockwise successive algorithm for a set $\mathcal{E}\left(i\right)$. }
\label{GBWS}
\end{figure}

Let $\mathbf{y}$ be an LLR vector of length $n$.
Let $\mathbf{p}$ be a vector of the same length such that $\mathbf{p}_i$, $0 \leq i < n$, is a probability that $\mathbf{y}_i$ has the incorrect sign.
These probabilities are calculated as 
\begin{equation}
\mathbf{p}_i = \frac{e^{-\left|\mathbf{y}_i\right|}}{1 + e^{-\left|\mathbf{y}_i\right|}}.
\label{probFromLLRs}
\end{equation}
Denote by $\mathcal{E}\left(i\right)$, $i \in \left\{0, 1, \dots, 2n-3 \right\}$, sets of indices corresponding to different higher-rate constituent codes.
For each $\mathcal{E}\left(i\right)$, we calculate 
\begin{equation}
E\left(\mathcal{E}\left(i\right)\right) = \sum_{j \in \mathcal{E}\left(i\right)}\mathbf{p}_j,
\label{errorsMean}
\end{equation}
i.e., the expected number of errors in the noisy vector corresponding to the higher-rate constituent code.
The generalized version of the blockwise successive decoding algorithm chooses $\mathcal{E}\left(i_j \right)$, $0 \leq j \leq p - 1$, with the smallest $E\left(\mathcal{E}\left({i_j}\right)\right)$.
The higher-rate constituent codes defined by $\mathcal{E}\left(i_j\right)$ are decoded using a sub-optimal decoding algorithm, e.g., automorphism-based recursive decoding or permutation-based blockwise successive decoding.
The hard output of this algorithm is used to change signs of LLRs $\mathbf{y}_k$, $k \in \left\lbrace 0, 1, \dots, n - 1 \right\rbrace \setminus \mathcal{E}\left(i_j\right)$.
The second half of the LLR vector is decoded by another sub-optimal decoder.
Finally, the algorithm returns a codeword with the smallest correlation discrepancy.
The formal description of this algorithm is presented in Algorithm \ref{FastPBSDAlg}.
Note that decoders of constituent codes are passed as arguments of the \texttt{GBWSDec} decoding function.
We denote decoders of the higher-rate and lower-rate constituent codes as $\texttt{uDec}$ and $\texttt{vDec}$, respectively.
Decoding for a set $\mathcal{E}\left(i\right)$ is schematically depicted in Fig. \ref{GBWS}.

In contrast to Algorithm \ref{PBSDAlg}, Algorithm \ref{FastPBSDAlg} computes $E\left(\mathcal{E}\left(i\right)\right)$, $0 \leq i < 2n - 2$, and selects $\mathcal{E}\left(i_0 \right), \mathcal{E}\left(i_1 \right), \dots \mathcal{E}\left(i_{p-1} \right)$ with the smallest expected number of errors.
It is done in $\mathcal{O}\left(n \log n\right)$.
Indeed, consider the LLR vector $\mathbf{y}$ of length $n=2^m$.
The vector $\mathbf{y}$ and \eqref{probFromLLRs} are used to calculate the vector of probabilities $\mathbf{p}$.
The result of FHT applied to the vector $\mathbf{p}$ can be written as $\mathbf{w} = \mathbf{p}\mathbf{H}$, where
\begin{equation*}
\mathbf{H} = \begin{bmatrix}
     1 &  1 \\
     1 & -1 \\
\end{bmatrix}^{\otimes m},
\end{equation*}
$\mathbf{X}^{\otimes m}$ denotes $m$-times Kronecker product of the matrix $\mathbf{X}$ with itself.
Note that any codeword of the first-order RM code $\mathcal{R}\left(1,m\right)$ can be written as $\left(\mathbf{1} \pm \mathbf{h}\right)/2$, where $\mathbf{1}$ is the all-ones vector of size $n$ and $\mathbf{h}$ is a column of the matrix $\mathbf{H}$.
Consequently, \eqref{errorsMean} can be calculated using the result of the FHT as $\left(\mathbf{w}_0 \pm \mathbf{w}_i\right)/2$, $0 < i < n$.
Since the running time of FHT and sorting is $\mathcal{O}\left(n \log n\right)$, $\mathcal{E}\left(i_0 \right), \mathcal{E}\left(i_1 \right), \dots \mathcal{E}\left(i_{p-1} \right)$ are found in $\mathcal{O}\left(n \log n\right)$.

\begin{algorithm}[!t]
\DontPrintSemicolon
\SetAlgoLined
\KwIn{An LLR vector $\mathbf{y}$ of length $n = 2^m$, a number of decompositions into shorter length RM codes $p$, constituent decoders $\texttt{uDec}$ and $\texttt{vDec}$}
\KwOut{A decoded codeword $\mathbf{c}$}
Let $\mathbf{c}$ be zero vector of length $n$, $M \gets \infty$ \;
Let $\mathcal{E}\left(i_0\right), \mathcal{E}\left(i_1\right), \dots \mathcal{E}\left(i_{p-1}\right) $ be $p$ sets of indices of higher-rate constituent codes with the smallest expected number of errors \eqref{errorsMean} \;

\For{$i = i_0, i_1, \dots i_{p - 1}$} {
	Let $\mathbf{h}$ be a vector of length $n$; $\mathbf{h}_j = 0$, if $j \in \mathcal{E}\left(i\right)$, otherwise $\mathbf{h}_j = 1$ \;
	Denote by $j_0, j_1, \dots j_{n/2-1}$ the indices such that $\mathbf{h}_{j_t} = 0$, $j_0 < j_1 < \dots < j_{n/2-1}$\;	
	Denote by $l_0, l_1, \dots l_{n/2-1}$ the indices such that $\mathbf{h}_{l_t} = 1$, $l_0 < l_1 < \dots < l_{n/2-1}$\;
	$\mathbf{y}^\prime \gets \left(\mathbf{y}_{j_0}, \mathbf{y}_{j_1}, \dots,  \mathbf{y}_{j_{n/2-1}} \right)$ \;
$\mathbf{y}^{\prime\prime} \gets \left(\mathbf{y}_{l_0}, \mathbf{y}_{l_1}, \dots,  \mathbf{y}_{l_{n/2-1}} \right)$ \;	
	$\mathbf{c}^\prime \gets \texttt{uDec}\left(\mathbf{y}^\prime \right)$ \;
	$\mathbf{y}^{\prime\prime}_t \gets \left(1 - 2\mathbf{c}^\prime_t \right)\mathbf{y}^{\prime\prime}_t$, for $t = 0, 1, \dots, n/2 - 1$\;
	$\mathbf{c}^{\prime\prime} \gets \texttt{vDec}\left(\mathbf{y}^{\prime\prime} \right)$ \;
   Let $\mathbf{\bar{c}}$ be zero vector of size $n$ \;
  $\mathbf{\bar{c}}_{j_t} \gets \mathbf{c}^\prime_t$, for $t = 0, 1, \dots, n/2 - 1$\; 
  $\mathbf{\bar{c}}_{l_t} \gets \mathbf{c}^\prime_t \oplus \mathbf{c}^{\prime\prime}_t$, for $t = 0, 1, \dots, n/2 - 1$\; 
	Let $\bar{M}$ be the correlation discrepancy of the codeword $\mathbf{\bar{c}}$\;
\lIf {$\bar{M} < M$} {$\mathbf{c} \gets \mathbf{\bar{c}}$, $M \gets \bar{M}$ }
}
 \Return{$\mathbf{c}$}
 \caption{The \texttt{GBWSDec} decoding function}
 \label{FastPBSDAlg}
\end{algorithm}

The running time of the \texttt{GBWSDec} function is roughly $p$ times the running time of constituent decoders $\texttt{uDec}$ and $\texttt{vDec}$ plus the running time of FHT and sorting.
Since the result of FHT is used to find indices of sets with the smallest expected number of errors, the algorithm needs to store the sets themselves.
Therefore, line 2 of Algorithm \ref{FastPBSDAlg} takes $\mathcal{O}\left(n^2\right)$ space.
It follows that the space complexity of Algorithm \ref{FastPBSDAlg} is at least $\mathcal{O}\left(n^2\right)$.
Note that the space complexity of Algorithm \ref{FastPBSDAlg} also depends on the space requirements of constituent decoders.

\begin{remark}
In \cite{PermGross2}, a permutation selection method has been proposed for recursive list decoding.
The key idea of the approach proposed in \cite{PermGross2} is to select the projected code maximizing the sum of LLR absolute values.
However, since a brute-force search is used to find the best projected code, the computational complexity of this approach is quite high.
To reduce the running time, the authors in \cite{PermGross2} proposed to perform the search through a small fraction of permutations called factor-graph permutations.
Note that this idea is used only for projected codes that do not allow for low-complexity ML decoding.
In contrast to \cite{PermGross2}, we consider a different decoding algorithm.
Although the performance of our decoding algorithm is limited, it allows for the low-complexity decomposition selection scheme that improves the error-rate performance significantly.
Furthermore, unlike the approach in \cite{PermGross2}, our scheme performs the search through all available decompositions.
\end{remark}

\section{Simulation Results}
In this section, we present simulation results for a BI-AWGN channel and compare the block error rate (BLER) performance with that of automorphism-based recursive decoding with $p$ permutations (referred to as AutRec-$p$).
In \cite{RMAutDec}, it has been demonstrated that the automorphism-based successive cancellation (SC) decoder outperforms the recursive list decoder with permutations \cite{Dumer} both in terms of error-rate performance and complexity.
To further improve the error-rate performance of automorphism-based decoding for the given number of permutations, we consider  the recursive decoder \cite{recDec} instead of the SC decoder.
In contrast to SC decoding in which the recursion is continued until codes of length 1, the recursive algorithm continues a decomposition procedure until an RM code allowing for low-complexity ML decoding, leading to an improvement in error-rate performance \cite{recDec}.
Specifically, we consider a version of recursive decoding that continues a decomposition until a first-order RM code or a parity-check code, i.e., $\mathcal{R}\left( h-1, h\right)$ for some $h$.
Furthermore, we use a "min-sum" approximation to decrease the running time of recursive decoding \cite{Bossert}.
The decoding algorithms considered in this section are briefly introduced in Table \ref{decodersBriefDescription}.

\begin{figure}[tbp]
\centering
\includegraphics[scale=0.98]{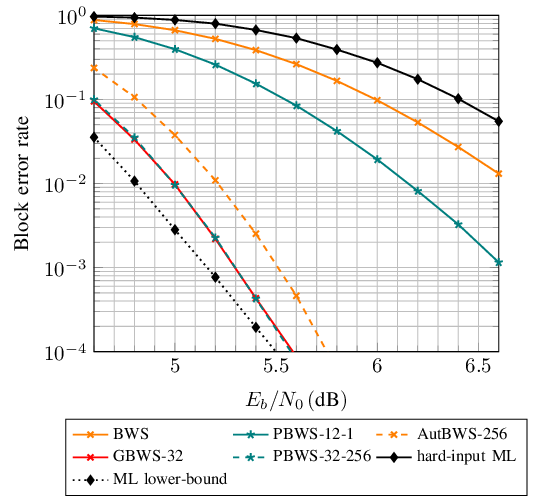}
\caption{The block error rate performance of $\mathcal{R}\left(7, 10\right)$ under different versions of blockwise successive decoding. For generalized blockwise successive decoding, the Chase decoder with 7 unreliable bits is used as \texttt{uDec} and PBWS-28-8 is used as \texttt{vDec}.}
\label{SDComp}
\end{figure}

In Fig. \ref{SDComp}, we present the performance of $\mathcal{R}\left(7, 10 \right)$ under different versions of blockwise successive decoders.
We denote the blockwise successive decoder by BWS, the permutation-based blockwise successive decoder with $l$ unreliable bits and $p$ permutations  as PBWS-$l$-$p$, and the generalized blockwise successive decoder with $p$ different decompositions as GBWS-$p$.
Note that, unless specified otherwise, the number of unreliable bits used in Chase decoding is upper bounded by 7.
It allows decreasing the running time of the proposed decoders at the cost of a negligible performance loss.

\begin{table*}[t]
\caption{Brief description of the decoding algorithms being compared.}

\begin{center}
\def\arraystretch{1.25}
\begin{tabular}{l|c|c}
\bfseries Decoding algorithm & \bfseries Constituent code decoded first & \bfseries Permutation selection \\
\hline
Automorphism-based recursive decoding (AutRec) & Lower-rate & Random \\
\hline
Blockwise successive decoding (BWS) & Higher-rate & -- \\ 
\hline
Permutation-based blockwise successive decoding (PBWS) & Higher-rate & Reliability-based (see Algorithm \ref{PermTransform}) \\
\hline
\begin{tabular}{@{}l@{}}Permutation-based blockwise successive decoding \\ with random permutations (AutBWS)
\end{tabular} & Higher-rate & Random \\
\hline
Generalized blockwise successive decoding (GBWS) & Higher-rate & Reliability-based using FHT (see Algorithm \ref{GBWS})\\
\hline
\begin{tabular}{@{}l@{}}
Generalized blockwise successive decoding \\ with random decompositions (RGBWS)
\end{tabular}  & Higher-rate & Random \\
\end{tabular}

\end{center}
\label{decodersBriefDescription}
\end{table*}

As expected, the blockwise successive decoding algorithm has very poor performance.
If the $\texttt{PermTransform}$ function is used to find a permutation for the received LLR vector, then the blockwise successive decoding performance is improved by more than 0.4 dB.
A larger number of permutations further improves the performance.
We can see that usage of random permutations causes degradation of 0.2 dB at a BLER of $10^{-3}$ compared to the decoder that uses soft information from a channel to generate permutations.
We denote the blockwise successive decoder with $p$ random permutations by AutBWS-$p$.
The permutation-based and generalized blockwise successive decoders have similar decoding error probability.
As will be shown later, these decoders also have similar complexity.
In Fig. \ref{SDComp}, we also plot the error-rate performance of a low-complexity hard-input ML decoding algorithm proposed recently \cite{Thangaraj} and the ML performance lower bound.
The ML lower bound is estimated using an approach in \cite{Dumer}.
Namely, we count the number of cases when the codeword returned by the generalized blockwise successive algorithm is more probable than the transmitted one, and plot the fraction of such events.
We observe that considered permutation-based and generalized decoders perform 0.1 dB from ML lower bound at a BLER of $10^{-4}$, while the hard-input ML decoder is not competitive with the soft-input algorithms.

\begin{figure*}[tb]

\centering
  \subfloat[$\mathcal{R}\left(5, 8\right)$, PBWS-$l$-32]{%
  	   \includegraphics[scale=1.05]{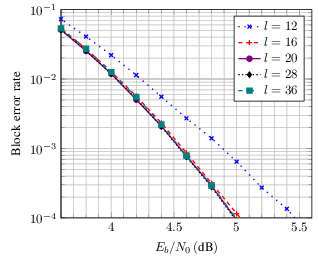}
       }
    \hfill
  \subfloat[$\mathcal{R}\left(6, 9\right)$, PBWS-$l$-64]{%
  		\includegraphics[scale=1.05]{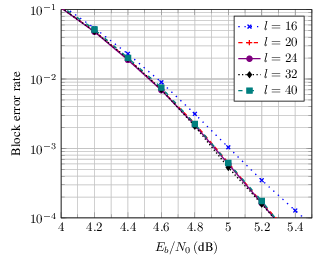}
        }
   \hfill
  \subfloat[$\mathcal{R}\left(7, 10\right)$, PBWS-$l$-256]{%
   		\includegraphics[scale=1.05]{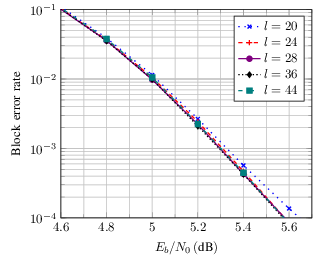}
        }
       \caption{The block error rate performance of $\mathcal{R}\left(m-3, m\right)$, $m \in \left\{8, 9 ,10\right\}$, under permutation-based blockwise successive decoding with different numbers of unreliable bits.}
	\label{lStudy}
\end{figure*}

\begin{figure*}[tb]

\centering
  \subfloat[$\mathcal{R}\left(5, 8\right)$]{%
  	   \includegraphics[scale=1.05]{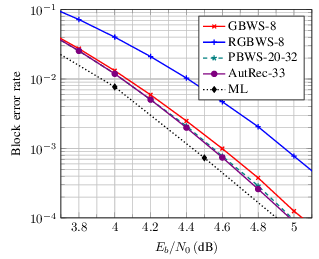}
       }
    \hfill
  \subfloat[$\mathcal{R}\left(6, 9\right)$]{%
  		\includegraphics[scale=1.05]{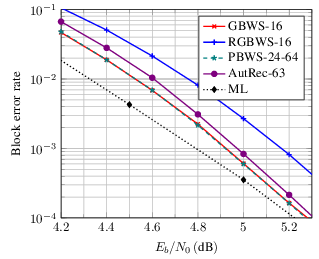}
        }
   \hfill
  \subfloat[$\mathcal{R}\left(7, 10\right)$\label{m_3_mc}]{%
  		\includegraphics[scale=1.05]{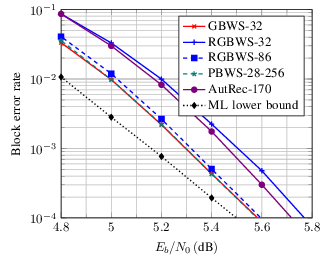}
        }
       \caption{The block error rate performance of $\mathcal{R}\left(m-3, m\right)$, $m \in \left\{8, 9 ,10\right\}$. For generalized blockwise successive decoding, the Chase decoder with 7 unreliable bits is used as $\texttt{uDec}$, while PBWS-20-4, PBWS-24-4, and PBWS-28-8  are used as $\texttt{vDec}$ for codes of length 256, 512, and 1024, respectively. RGBWS-$p$ denotes a version of generalized blockwise successive decoding that chooses $p$ decompositions into shorter codes at random. ML simulation results are taken from \cite{ML}.}
	\label{m_3_m}
\end{figure*}

\begin{figure*}[t]

\centering
  \subfloat[$\mathcal{R}\left(3, 7\right)$]{%
  	   \includegraphics[scale=1.05]{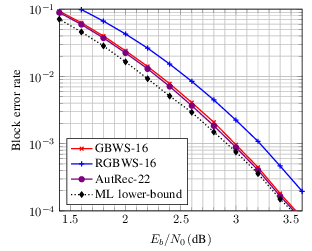}
       }
    \hfill
  \subfloat[$\mathcal{R}\left(3, 8\right)$\label{simResb}]{%
  		\includegraphics[scale=1.05]{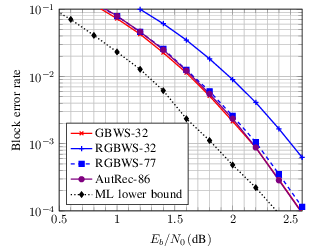}
        }
   \hfill
  \subfloat[$\mathcal{R}\left(4, 8\right)$]{%
  		\includegraphics[scale=1.05]{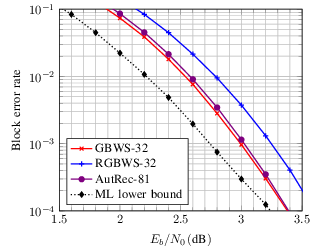}
        }
    \\
  \subfloat[$\mathcal{R}\left(3, 9\right)$]{%
  	   \includegraphics[scale=1.05]{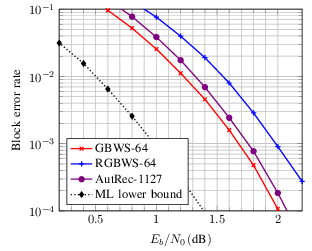}
       }
    \hfill
  \subfloat[$\mathcal{R}\left(4, 9\right)$\label{simRese}]{%
  		\includegraphics[scale=1.05]{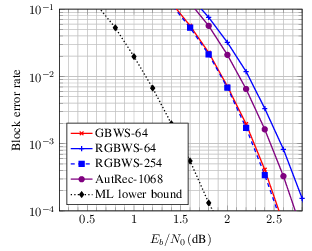}
        }
   \hfill
  \subfloat[$\mathcal{R}\left(5, 9\right)$]{%
  		\includegraphics[scale=1.05]{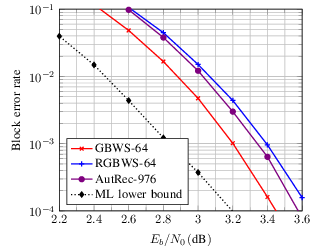}
        }
\caption{The block error rate performance of RM codes. For generalized blockwise successive decoding, AutRec-2, AutRec-4, and AutRec-32 are used as $\texttt{uDec}$, while recursive decoding, AutRec-2, and AutRec-8 are used as $\texttt{vDec}$ for codes of length 128, 256, and 512, respectively. RGBWS-$p$ denotes a version of generalized blockwise successive decoding that chooses $p$ decompositions into shorter codes at random.}
\label{simRes}
\end{figure*}

In Fig. \ref{lStudy}, we present the error-rate performance of the permutation-based blockwise successive decoder with different numbers of unreliable bits.
We see that, for a wide range of the number of unreliable bits, the permutation-based blockwise successive decoder demonstrates almost the same performance.
Thus, careful optimization of this parameter does not allow for better error-rate performance.

In Figs. \ref{m_3_m} and \ref{simRes}, we compare the error-rate performance of the proposed decoders to that of automorphism-based recursive decoding.
For decoders considered in Figs. \ref{m_3_m} and \ref{simRes}, we estimate the complexity by counting the number of floating-point operations (comparisons and additions/subtractions) required to decode one codeword.
Since the number of operations depends on the level of noise in the channel, we report complexity for an $E_b/N_0$ of $-10$ dB in Table \ref{runingTimeComparisonM10} and for the $E_b/N_0$ required to achieve a BLER of $10^{-4}$ in Table \ref{runingTimeComparisonEM4}.
For automorphism-based recursive decoding, this difference is primarily caused by the decoder of parity-check constituent codes, because, if the parity-check is satisfied, then there is no need to search for the minimum absolute value.
Although the complexity of automorphism-based recursive decoding is roughly the same for both scenarios, it is significantly improved for the considered blockwise successive decoders in the high signal-to-noise ratio (SNR) region in the case of $\mathcal{R}\left(m-3, m\right)$, $m \in \left\{8, 9 ,10\right\}$.
This improvement is due to Chase decoding used in the proposed algorithms.
The Chase decoder calculates the syndrome and, if it is the all-zero vector, then the output codeword is immediately obtained by taking signs of LLRs.
Thus, the average-case complexity of the Chase algorithm is improved in the high SNR region.
In our simulations, we compare the generalized blockwise successive decoding to automorphism-based decoding with nearly the same complexity in the low SNR region.

\begin{table*}[t]
\caption{Number of floating-point operations required to decode one codeword at an $E_b/N_0$ of $-10\:\mathrm{dB}$.}
\begin{center}
\begin{tabular}{|c|c|c|c|c|c|c|c|c|c|}
\hline
Code & $\mathcal{R}\left(3,7\right)$ & $\mathcal{R}\left(3,8\right)$ & $\mathcal{R}\left(4,8\right)$ & $\mathcal{R}\left(5,8\right)$ & $\mathcal{R}\left(3,9\right)$ & $\mathcal{R}\left(4,9\right)$ & $\mathcal{R}\left(5,9\right)$ &$\mathcal{R}\left(6,9\right)$ & $\mathcal{R}\left(7,10\right)$ \\
\hline
PBWS & -- & -- & -- & 46361 & -- & -- & -- & 147004 & 867311\\
\hline
GBWS & 19213 & 186173 & 142802 & 39943 & 5699679 & 4775932 & 3427892 & 148814 & 797169 \\
\hline
AutRec & 19642 & 187752 & 144142 & 40372 & 5702795 & 4777835 & 3429521 & 150676 & 801220\\
\hline
\end{tabular}

\end{center}
\label{runingTimeComparisonM10}
\end{table*}

\begin{table*}[!t]
\caption{Number of floating-point operations required to decode one codeword at a BLER of $10^{-4}$.}

\begin{center}
\begin{tabular}{|c|c|c|c|c|c|c|c|c|c|}
\hline
Code & $\mathcal{R}\left(3,7\right)$ & $\mathcal{R}\left(3,8\right)$ & $\mathcal{R}\left(4,8\right)$ & $\mathcal{R}\left(5,8\right)$ & $\mathcal{R}\left(3,9\right)$ & $\mathcal{R}\left(4,9\right)$ & $\mathcal{R}\left(5,9\right)$ &$\mathcal{R}\left(6,9\right)$ & $\mathcal{R}\left(7,10\right)$ \\
\hline
PBWS & -- & -- & -- & 26078 & -- & -- & -- & 85475 & 566786\\
\hline
GBWS & 18314 & 181253 & 134929 & 25740 & 5646313 & 4687249 & 3268816 & 101404 & 565371 \\
\hline
AutRec & 18882 & 183789 & 137768 & 36620 & 5653858 & 4690475 & 3289294 & 135938 & 722947\\
\hline
\end{tabular}

\end{center}
\label{runingTimeComparisonEM4}
\end{table*}

\begin{figure*}[t]

\centering
  \subfloat[$\mathcal{R}\left(3, 8\right)$, $1.8$ dB]{%
  		
       \includegraphics[scale=1.05]{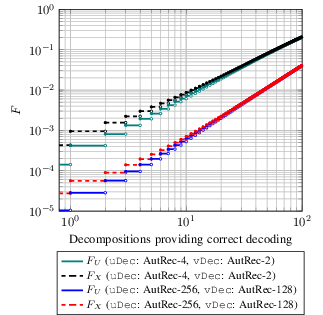}
       }
    \hfill
  \subfloat[$\mathcal{R}\left(4, 9\right)$, $1.6$ dB]{%
  		\includegraphics[scale=1.05]{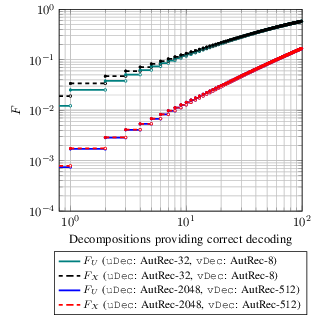}
        }
   \hfill
  \subfloat[$\mathcal{R}\left(7, 10\right)$, $5.2$ dB]{%
  		\includegraphics[scale=1.05]{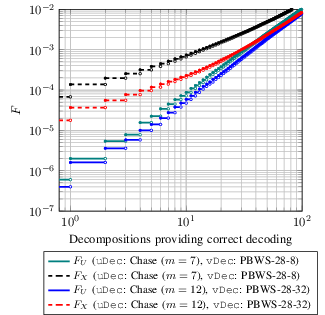}
        }
       \caption{Empirical cumulative distribution functions of the number of decompositions resulting in correct decoding.}
	\label{decStat}
\end{figure*}

In Fig. \ref{m_3_m}, we plot the BLER for $\mathcal{R}\left(m-3, m \right)$, $m \in \left\{8, 9, 10 \right\}$.
For the generalized blockwise successive decoder, we use the Chase decoding algorithm as $\texttt{uDec}$ and the permutation-based blockwise successive decoder as $\texttt{vDec}$.
The proposed decoders perform within $0.15$ dB from ML decoding.
In the case of $\mathcal{R}\left(5, 8 \right)$, the permutation-based blockwise successive decoder and the automorphism-based decoder perform similarly, while the proposed decoding algorithms offer better performance for $\mathcal{R}\left(6, 9 \right)$ and $\mathcal{R}\left(7, 10 \right)$.
For instance, in the case of $\mathcal{R}\left(7, 10 \right)$, the generalized blockwise successive decoder outperforms the automorphism-based recursive decoder by 0.16 dB at a BLER of $10^{-3}$.
Furthermore, on average, the permutation-based and generalized blockwise successive decoders require at least 20\% fewer operations to decode one codeword at a BLER of $10^{-4}$.
To demonstrate the improvement due to the proposed constituent codes selection scheme, we report the error-rate performance of a decoder that randomly chooses $p$ decompositions into shorter codes (referred to as RGBWS-$p$).
Observe that the proposed selection technique gives a gain of at least $0.2$ dB over the randomized one.

In Fig. \ref{simRes}, we present BLER results for RM codes of various orders and lengths.
We also plot the ML performance lower bound obtained using graph search decoding \cite{GSD}.
For codes of length 128 and 256, the generalized blockwise successive decoder provides a performance close to that of automorphism-based recursive decoding.
The proposed decoding algorithm outperforms the automorphism-based recursive decoder for codes of length 512.
Specifically, at a BLER of $10^{-4}$, the generalized blockwise successive decoder shows a performance gain of 0.07 dB, 0.17 dB, and 0.13 dB for codes of order 3, 4, and 5, respectively.
As in Fig. \ref{m_3_m}, we can see that the proposed decomposition selection technique improves upon the randomized one.

Finally, for the generalized blockwise successive decoder, we investigate the number of decompositions leading to correct decoding.
Consider an RM code $\mathcal{R}\left(r, m \right)$ and assume that $\texttt{uDec}$ and $\texttt{vDec}$ algorithms are used for decoding of $\mathcal{R}\left(r, m-1 \right)$ and $\mathcal{R}\left(r-1, m-1 \right)$ constituent codes, respectively.
Recall that there are $2^{m+1}-2$ different decompositions of $\mathcal{R}\left(r, m \right)$ into $\mathcal{R}\left(r, m-1 \right)$ and $\mathcal{R}\left(r-1, m-1 \right)$.
Let the random variable $U$ denote the number of decompositions, in which a higher-rate constituent code is decoded correctly, and let the random variable $X$ denote the number of decompositions, in which both constituent codes are decoded correctly.
In Fig. \ref{decStat}, we plot the empirical cumulative distribution functions of the random variables $U$ and $X$.
Note that we consider SNRs, at which the experimental ML performance lower bound is approximately $10^{-3}$.

For $\mathcal{R}\left(3 ,8 \right)$ and $\mathcal{R}\left(4 ,9 \right)$, we clearly see that, for the given pair of constituent decoders, the probability that there are $s$ "good" decompositions is roughly the same for both $U$ and $X$.
Thus, if a higher-rate constituent code is decoded correctly, then $\texttt{vDec}$ returns the correct codeword with high probability.
Furthermore, if we use a more powerful algorithm for decoding of a higher-rate constituent code, then the probability that there is at least one "good" decomposition is increased.
In the case of $\mathcal{R}\left(7 , 10 \right)$, we use Chase decoding as the decoder of a higher-rate constituent code and we observe only a slight improvement upon the decoder with a smaller number of unreliable bits.
The reason for this is that Chase decoding even with 7 unreliable bits achieves near-ML performance for the considered case.

Given the empirical cumulative distribution function $F_X$ of the random variable  $X$, one can approximate the error-rate performance of the decoder with random decomposition selection.
Indeed, let us assume that decompositions are independent, i.e., the success of decoding for a given decomposition is independent of success for others, and let us assume that the transmitted codeword is the closest codeword to the received sequence, i.e., the ML decoder always returns the correct codeword.
Under these assumptions, the probability that $p$ random decompositions result in incorrect decoding equals
\begin{equation}
\begin{aligned}
\sum\limits_{i=0}^{2n - 2 - p} P\left(X = i\right)\frac{{2n-2 - i \choose p }{i \choose 0} }{{2n - 2 \choose p}} + \sum\limits_{i=2n - 2 - p + 1}^{2n - 2} P\left(X = i\right) \\
= \sum\limits_{i=0}^{2n - 2-p} P\left(X = i\right)\prod\limits_{j=0}^{p-1}\frac{2n-2-i-j}{2n-2-j} \\
 + \sum\limits_{i=2n - 2 - p + 1}^{2n - 2} P\left(X = i\right),
\end{aligned}
\label{probEst}
\end{equation}
where $n$ is the code length and $P\left(X = i\right) = F_X\left(i \right) -  F_X\left(i-1 \right)$.
In Figs. \ref{m_3_mc}, \ref{simResb}, and \ref{simRese}, we plot the BLER performance of the decoder that randomly chooses $p$ decompositions into shorter codes and performs close to the generalized blockwise successive decoder.
The number of decompositions $p$ is selected as the smallest value, for which \eqref{probEst} gives a lower probability of error compared to the generalized blockwise successive decoder.
We can see that the error-rate performance of considered decoders is nearly the same.

\section{Conclusion}
In this paper, we presented non-iterative soft-input decoding algorithms that, unlike traditional recursive decoding, start decoding with a constituent code of the same order.
Although the error-rate performance of the proposed blockwise successive decoder is limited, we showed that it can be significantly improved by means of a clever choice of permutations employing soft information from a channel.
For short length ($\leq 256$) RM codes, we demonstrated a performance close to automorphism-based recursive decoding with the same computational complexity, while, for longer RM codes, it is shown that the proposed decoders offer a performance gain.

For RM codes of length $2^m$ and order $m-3$, the proposed algorithms perform within 0.15 dB from ML decoding at a BLER of $10^{-4}$.
Furthermore, due to the use of the Chase II decoding algorithm, our decoders take a shorter average-case running time to decode one codeword compared to the automorphism-based decoder.
In particular, for $\mathcal{R}\left(7, 10 \right)$, we showed that the proposed algorithms outperform the automorphism-based recursive decoder by 0.13 dB at a BLER of $10^{-4}$, while, on average, requiring 21\% fewer operations to decode one codeword.

\bibliographystyle{IEEEtran}
\bibliography{IEEEabrv,recursiveDecodingStartingWithHigherRateCodeBib}

\end{document}